\documentclass[aps,prb,showpacs,showkeys,twocolumn,groupedaddress]{revtex4-1}  

\usepackage{amsmath,amssymb}
\usepackage[dvips]{graphicx}
\usepackage{color}
\begin{document}
\newcommand{\dd}{\mathrm{d}}
\newcommand{\ii}{\mathrm{i}}
\newcommand{\ee}{\mathrm{e}}
\newcommand{\Sp}{\mathrm{Sp}}
\newcommand{\heav}{{\text{\usefont{U}{psy}{m}{n}\selectfont\symbol{113}}}}
\newcommand{\dirac}{{\text{\usefont{U}{psy}{m}{n}\selectfont\symbol{100}}}}
\newcommand{\kronek}{{\text{\usefont{U}{psy}{m}{n}\selectfont\symbol{100}}}}
\newcommand{\lindhard}{\mathrm{L}}
\newcommand{\jell}{\mathrm{jell}}
\newcommand{\unif}{\mathrm{unif}}
\newcommand{\bulk}{\mathrm{bulk}}
\newcommand{\surf}{\mathrm{surf}}
\newcommand{\const}{\mathrm{const}}

\newcommand{\BLUE}{\textcolor{blue}}

\numberwithin{equation}{section}


\title{Semi-infinite jellium: Step potential model}


\author{P. P. Kostrobij, B. M. Markovych}
\email[]{bogdan\_markovych@yahoo.com}
\affiliation{Lviv Polytechnic National University, 12 Bandera Str., 79013 Lviv, Ukraine}


\date{\today}

\begin{abstract}
 The surface energy,
 the one-particle distribution function of electrons, etc.
 of a semi-bounded metal within the framework of the semi-infinite jellium are calculated.
 The influence of the potential barrier height on these characteristics is studied.
 The barrier height is found from the condition of the minimum of the surface energy.
 The surface energy is positive in the entire domain of the Wigner-Seitz radius of metals,
 and it is in sufficiently good agreement with experimental data.
\end{abstract}


\maketitle


 \section{Introduction}

 In Refs.\,\cite{KMpreprint2014, KMprb2015}, using the method of functional integration,
 the quantum-statistical theory of simple semi-bounded metal within the framework of
 the semi-infinite jellium is built.
 The advantage of this theory is taking into account of the Coulomb interaction between electrons
 in the semi-bounded system.
 In particular,
 using the infinite barrier model of the surface potential,
 the one-particle distribution function of electrons and
 the surface energy are calculated.
 This model potential is the simplest,
 but it is obtained an important result ---
 the surface energy of the semi-infinite jellium is positive
 in the entire area of the Wigner-Seitz radius of metals.
 Conversely, the usage of the density functional theory,
 which today is the most popular,
 leads to the well known  problem of the surface energy negative values at high concentrations of electrons.
 Overview of papers that focus on the study of the surface energy is in Refs.\,\cite{KMpreprint2014,KMprb2015}.

 This paper is a continuation of Refs.\,\cite{KMpreprint2014,KMprb2015},
 the difference is in the way of modeling the surface potential, namely,
 in using of the step potential model for the surface potential.
 Moreover the height of the potential barrier is found from the minimum of the surface energy.
 The calculated values of the surface energy are somewhat lower
 than obtained in Refs.\,\cite{KMpreprint2014, KMprb2015} for the infinite barrier model.
 These values are in sufficiently good agreement with experimental data.
 The calculation of the one-particle distribution function of electrons shows that
 this function slower goes down to zero out of the positive charge
 in comparison with the one-particle distribution function in the infinitely high potential barrier.
 It is shown that
 if the potential barrier height tends to infinity,
 the obtained results coincide with the results of Ref.\,\cite{KMpreprint2014, KMprb2015}.

 \section{Thermodynamic potential}

 We consider a semi-bounded metal within the framework of the semi-infinite jellium,
 i.e. a system of $N$ electrons is located in the volume $V=SL$ in the field of positive charge,
 which is bounded by the dividing plane ${Z=0}$,
 with the distribution
 \begin{align*}\label{PositiveCharge}
   \varrho_\jell({\bf R})&\equiv\varrho_\jell({\bf R}_{||},Z)\equiv\varrho_\jell(Z)\\
   &=\varrho_0\heav(-Z)=
    \left\{\begin{array}{ll}
                      \varrho_0, & Z\leqslant0, \\
                      0, & Z>0,
                    \end{array}
    \right.
 \end{align*}
 where $\heav(x)$ is the Heaviside function,
 ${\bf R}_{||}=(X,Y)$,
 $X,Y\in [-\sqrt{S}/2,+\sqrt{S}/2]$,
 $Z\in\left[-L/2,+L/2\right]$,
 moreover,
 the condition of electroneutrality is satisfied,
 \begin{equation}\label{electroNeutr}
  \lim_{S,L\to\infty}
  \int\limits_S \! \dd{\bf R}_{||} \!\!
  \int\limits_{-L/2}^{+L/2} \!\!\!\! \dd Z \,
  \varrho_\jell({\bf R}_{||},Z) = e N, \; e>0,
 \end{equation}
 and, in the thermodynamic limit, we have
 \begin{equation*}
    \lim_{N,S,L\to\infty}\frac{eN}{SL}=
    \lim_{N,V\to\infty}\frac{eN}{V/2}=
    \varrho_0.
 \end{equation*}

 The expression for the thermodynamic potential of this system is obtained in Refs.~\cite{KMpreprint2014,KMprb2015},
 \begin{equation*}
    \Omega=\Omega_0
           -\frac1{2S}\langle N\rangle_0\sum\limits_{\mathbf{q}\neq0}\nu(\mathbf{q},0)
           +\Omega_{\mathrm{int}},
 \end{equation*}
 where
 \begin{equation}\label{omegaIdeal0}
    \Omega_0=-\frac1\beta\sum\limits_{\mathbf{k}_{||},\alpha}
    \ln\left[1+\ee^{\beta(\mu-E_\alpha(\mathbf{k}_{||}))}\right]
 \end{equation}
 is the thermodynamic potential of the noninteracting system\cite{footnote},
 $\beta=1/\theta$, $\theta$ is the thermodynamic temperature,
 $\mu=\frac{\hbar^2\mathcal{K}^2_\mathrm{F}}{2m}$
 is the chemical potential of the system taking into account the Coulomb interaction between electrons,
 $\mathcal{K}_\mathrm{F}$ is the magnitude of the Fermi wave vector,
 $E_\alpha(\mathbf{k}_{||})=\frac{\hbar^2k^2_{||}}{2m}+\varepsilon_\alpha$
 is the energy of the electron in the field of the surface potential $V_\surf(z)$,
 in the state $(\mathbf{k}_{||},\alpha)$,
 $\mathbf{k}_{||}$ is the wave vector of the electron in the plane parallel to the dividing plane,
 $\alpha$ is a quantum number that depends on the form of the surface potential;
 \begin{equation}\label{N0}
  \langle N\rangle_0=\sum\limits_{\mathbf{k}_{||},\alpha}n_\alpha(\mathbf{k}_{||})
 \end{equation}
 is the average of the number operator of electrons
 (averaging is performed without taking into account the Coulomb interaction between electrons~\cite{KMpreprint2014,KMprb2015}),
 \begin{equation*}
    n_\alpha(\mathbf{k}_{||})=\frac1{\ee^{\beta(E_\alpha(\mathbf{k}_{||})-\mu)}+1}
 \end{equation*}
 is the Fermi-Dirac distribution,
  $\nu(\mathbf{q},z)=\frac{2\pi e^2}{q}\, \ee^{-q|z|}$
 is the two-dimensional Fourier transform of the Coulomb interaction,
 ${\mathbf{q}=(q_x,q_y)}$,
 ${q_{x,y}=\frac{2\pi}{\sqrt{S}}m_{x,y}}$,
 ${m_{x,y}=0,\pm1,\pm2,\ldots}$,
 $z$ is the electron coordinate normal to the dividing plane;
 \begin{align*}
    \Omega_{\mathrm{int}}&=
     -\frac1{2SL^2}\sum\limits_{\mathbf{q}\neq0,\nu}
     \int\limits_{-\frac L2}^{+\frac L2}\!\!\dd z_1\!\!
     \int\limits_{-\frac L2}^{+\frac L2}\!\!\dd z_2\;
     D(\mathbf{q},\nu,z_1,z_2)\\
     &\quad\times\int\limits_0^1\!\!
     g(\mathbf{q},\nu,z_1,z_2,\lambda)
     \dd\lambda,
 \end{align*}
 where $D(\mathbf{q},\nu,z_1,z_2)$ is the effective two-particle correlator
 taking into account the Coulomb interaction between electrons~\cite{KMpreprint2014,KMprb2015},
 $\nu$ is Bose frequency,
 $g(\mathbf{q},\nu,z_1,z_2,\lambda)$ is the effective interelectron interaction potential
 in $(\mathbf{q},z)$ representation,
 which depends on the parameter~$\lambda$
 and is a solution of the integral equation~\cite{KMpreprint2014,KMprb2015}
 \begin{align}\label{intEq1}
   g(\mathbf{q},&\nu,z_1,z_2,\lambda)=\nu({\bf q},z_1-z_2) \nonumber\\
                &+\frac\beta{S L^2}\,\lambda\!\!
                   \int\limits_{-\frac L2}^{+\frac L2} \!\! \dd z \!\!
                   \int\limits_{-\frac L2}^{+\frac L2} \!\! \dd z' \,
                   \nu({\bf q}|z_1-z')D(\mathbf{q},\nu,z',z)\nonumber\\
                &\quad\times   g(\mathbf{q},\nu,z,z_2,\lambda).
 \end{align}

 In Refs. \cite{KMpreprint2014,KMprb2015},
 it is shown that in the random phase approximation
 and neglecting the dependence of the effective interelectron interaction on Bose frequency~$\nu$
 the thermodynamic potential has the form
 \begin{align}\label{omega1}
    \Omega&=\Omega_0+\frac1{2S}\sum\limits_{\mathbf{q}\neq0}
                     \sum\limits_{\mathbf{k}_{||},\alpha}n_\alpha(\mathbf{k}_{||})
      \int\limits_{-\frac L2}^{+\frac L2}\!\!\dd z\,|\varphi_\alpha(z)|^2 \nonumber \\
      &\qquad\times
      \int\limits_0^1\!\!\dd\lambda \,\big(g(\mathbf{q},z,z,\lambda)-\nu(\mathbf{q},0)\big)\nonumber\\
    &\quad-\frac1{2S}\sum\limits_{\mathbf{q}\neq0}
                     \sum\limits_{\mathbf{k}_{||},\alpha_1,\alpha_2}
      n_{\alpha_1}(\mathbf{k}_{||}) \,
      n_{\alpha_2}(\mathbf{k}_{||}-\mathbf{q}) \nonumber\\
    &\qquad\times\int\limits_{-\frac L2}^{+\frac L2}\!\!\dd z_1\!\!
      \int\limits_{-\frac L2}^{+\frac L2}\!\!\dd z_2\,
      \varphi^*_{\alpha_1}\!(z_1) \varphi^{\vphantom{*}}_{\alpha_2}\!(z_1)
      \varphi^*_{\alpha_2}\!(z_2) \varphi^{\vphantom{*}}_{\alpha_1}\!(z_2) \nonumber\\
    &\qquad\times
      \int\limits_0^1\!\!\dd\lambda \,g(\mathbf{q},z_1,z_2,\lambda),
 \end{align}
 where the functions $\varphi_\alpha(z)$ satisfy the one-dimensional stationary Schr\"odinger equation with the surface potential $V_\surf(z)$~\cite{KMpreprint2014,KMprb2015}
 \begin{equation}\label{SchredEq}
   \left[-\frac{\hbar^2}{2m}\frac{\dd^2}{\dd z^2}+V_\surf(z)\right] \varphi_\alpha(z)=\varepsilon_\alpha\varphi_\alpha(z).
 \end{equation}

 \section{Step potential model}\label{FBM}

 In this work,
 the surface potential is modeled by the step potential
 of the height ${W=\frac{\hbar^2s^2}{2m}=\widetilde{s}^2\mu}$,
 where $\widetilde{s}$ is the barrier height parameter,
 which determines the barrier height,
 i.e.
 \begin{equation}\label{sTilda}
  \widetilde{s}=s/\mathcal{K}_\mathrm{F}=\sqrt{W/\mu},
 \end{equation}
 which is placed at the point $z=d$, i.e.
 \begin{equation}\label{stepPot}
  V_\surf(z)=
  \left\{
    \begin{array}{cl}
      W, & z>d, \\
      0, & z<d,
    \end{array}
  \right.
 \end{equation}
 and allows analytical solution of the one-dimensional stationary Schr\"odinger equation~\eqref{SchredEq}.
 Such solution, that satisfies the boundary conditions,
 \begin{equation*}
   \varphi_\alpha(-L/2)=0,\quad
   \varphi_\alpha(+\infty)=0,
 \end{equation*}
 the conditions of continuity and smoothness,
 \begin{equation*}
  \left\{
   \begin{array}{c}
     \varphi_\alpha(z<d)\big|_{z=d}=\varphi_\alpha(z>d)\big|_{z=d}, \\[2mm]
     \varphi_\alpha'(z<d)\big|_{z=d}=\varphi_\alpha'(z>d)\big|_{z=d}
   \end{array}
  \right.
 \end{equation*}
 is
 \begin{equation}\label{waveFunction}
  \begin{split}
        \varepsilon_\alpha&=\frac{\hbar^2\alpha^2}{2m},\\
        \varphi_\alpha(z)&=C(\alpha)
    \left\{
     \begin{array}{ll}
       \sin\big(\alpha(d-z)+\gamma(\alpha)\big), & z\leqslant d, \\[2mm]
       \dfrac{\alpha}{s} \,\ee^{-\varkappa(\alpha)(z-d)}, & z>d,
     \end{array}
    \right. \\
  \end{split}
 \end{equation}
 where
 \begin{align*}
  \gamma(\alpha)&=\arcsin\frac{\alpha}{s},\\
  \varkappa(\alpha)&=\left(\frac{\dd\gamma(\alpha)}{\dd\alpha}\right)^{-1}\!\!\!\!\!\!=\sqrt{s^2-\alpha^2},\quad
  \alpha\leqslant s,
 \end{align*}
 and quantum numbers $\alpha$ satisfy the algebraic transcendental equation,
 \begin{equation}\label{eqForAlfa}
  \alpha\left(\frac{L}{2}+d\right)+\gamma(\alpha)=\pi n,\;
  n=1,2,3,\ldots
 \end{equation}

 From the normalization condition for the wave functions,
 \[
   \int\limits_{-\frac L2}^{+\infty}\!\!
   |\varphi_\alpha(z)|^2\dd z=1,
 \]
 it follows that
 \begin{align*}
      C(\alpha)&=\frac{2}{\sqrt{L+2\left(d+\frac1{\varkappa(\alpha)}\right)}}\\
  &=\frac{2}{\sqrt{L+2\left(d+\frac{\dd\gamma(\alpha)}{\dd\alpha}\right)}}  .
 \end{align*}


 Note that the electron states ${\varepsilon_\alpha>W}$ are not written out,
 because only the states ${\varepsilon_\alpha\leqslant \mu}$ are really interesting for us,
 and for physically interesting problems
 the chemical potential of electrons is less than the barrier height,
 ${\mu\leqslant W}$.

 The step potential~\eqref{stepPot} has the parameter $d$,
 which determines the position of the potential barrier,
 and is determined by the condition of electroneutrality~\eqref{electroNeutr}.
 It is necessary to calculate the one-particle distribution function of electrons.

 \section{One-particle distribution function of electrons}\label{unarnaFunc}

 Let us calculate the one-particle distribution function of electrons~\cite{JPS2003_2}
 for the step potential model~\eqref{stepPot}
 in the case of low temperatures
 \begin{align*}
    F_1^0(z)=&\frac{V}{\langle N\rangle_0}\frac1S
              \sum\limits_{\mathbf{k}_{||},\alpha}
              |\varphi_\alpha(z)|^2n_\alpha(\mathbf{k}_{||})\\
            =&\frac{V}{\langle N\rangle_0}\frac1S
              \sum\limits_{\mathbf{k}_{||},\alpha}
              |\varphi_\alpha(z)|^2\heav\big(\mathcal{K}_{\mathrm{F}}^2-k_{||}^2-\alpha^2\big).
 \end{align*}
 Transition from the sums to the integrals
 according to rules~\cite{KMpreprint2014,KMprb2015},
 \begin{equation*}
    \sum\limits_{\mathbf{k}_{||}}f({k}_{||})=
    \frac{2S}{(2\pi)^2}
    \int\limits_{-\infty}^{+\infty}\!\!\!
    \dd \mathbf{k}_{||}\,f({k}_{||})=
    \frac{S}{\pi}\int\limits_0^\infty\!\!
    \dd k_{||}\,k_{||}\,f(k_{||}),
 \end{equation*}
 \begin{align}\label{SumToIntAlpha}
    \sum\limits_{\alpha}f(\alpha)&=
    \int\limits_{0}^{\infty}\!\!
    \dd\alpha
    \Bigg[\frac{L}{2\pi}\left(1+\frac2L\left(d+\frac{\dd\gamma(\alpha)}{\dd\alpha}\right)\right)\nonumber\\
    &\qquad-\frac12\dirac(\alpha)\Bigg] f(\alpha)=\nonumber\\
    &=\int\limits_{0}^{\infty}\!\!
    \dd\alpha
    \left(
     \frac{2}{\pi|C(\alpha)|^2}
     -\frac12\dirac(\alpha)
    \right)
    f(\alpha),
 \end{align}
 and integration with respect to the variable~$k_{||}$ lead to
 \begin{align}\label{unarna}
    F_1^0(z)&=\frac{3}{\mathcal{K}_\mathrm{F}^3}
    \int\limits_0^{\mathcal{K}_\mathrm{F}}\!\!
    \dd\alpha\, (\mathcal{K}_\mathrm{F}^2-\alpha^2)\nonumber\\
    &\quad\times\left\{
    \begin{array}{ll}
      \displaystyle\sin^2\big(\alpha(d-z)+\gamma(\alpha)\big), & z\leqslant d, \\
      \displaystyle\frac{\alpha^2}{s^2}\,\ee^{-2\varkappa(\alpha)(z-d)}, & z>d.
    \end{array}
    \right.
 \end{align}
 Integration with respect to the variable $\alpha$ must be performed numerically.

 If in Eq.~\eqref{unarna} the barrier height tends to infinity,
 this equation takes the well-known form~\cite{KMpreprint2014, KMprb2015}
  \begin{align}\label{unarnaIBM}
  F_1^0(z)&=
  \left[
   1+\frac{3\cos\big(2\mathcal{K}_{\mathrm{F}}(d-z)\big)}{\big(2\mathcal{K}_{\mathrm{F}}(d-z)\big)^2}
   -\frac{3\sin\big(2\mathcal{K}_{\mathrm{F}}(d-z)\big)}{\big(2\mathcal{K}_{\mathrm{F}}(d-z)\big)^3}
  \right]\nonumber\\
  &\quad\times\heav(d-z),
 \end{align}
 which is the one-particle distribution function of electrons in the case of
 infinite potential barrier model.

 In Fig.\,\ref{F01rs=2},
 the one-particle distribution function of electrons~\eqref{unarna}
 as a function of the electron coordinate normal is presented for
 the following values of Wigner-Seitz radius:
 $r_\mathrm{s}=2\,a_\mathrm{B}$ and $r_\mathrm{s}=6\,a_\mathrm{B}$,
 and different values of the barrier height parameter.
 The solid line represents the one-particle distribution function of electrons,
 which depends on the chemical potential of interacting electrons.
 The dashed line represents the one-particle distribution function of electrons
 without the Coulomb interaction.
 The positive charge is located in the domain ${z\leqslant0}$.
 It can be concluded:
 (1) taking into account the Coulomb interaction leads to
 an increase of the period of damping oscillations of
 the one-particle distribution function around its value
 in the bulk of the metal, which equals to unity;
 and (2) increasing of the barrier height leads to more rapid damping
 of the one-particle distribution function near the dividing plane.

 The parameter $d$ is determined by the condition of electroneutrality~\eqref{electroNeutr},
 which for the one-particle function has the form
 \[
  \lim_{L\to\infty}
  \int\limits_{-\frac{L}{2}}^{+\frac{L}{2}}\!\!
  \big(
   F_1^0(z)-\heav(-z)
  \big)
  \dd z=0.
 \]
 From this condition it follows that
 \begin{equation*}
    d=\frac{3\pi}{8\mathcal{K}_\mathrm{F}}
     -\frac{3}{2\mathcal{K}_\mathrm{F}^3}
     \int\limits_0^{\mathcal{K}_\mathrm{F}}\!\!
    \dd\alpha\, (\mathcal{K}_\mathrm{F}^2-\alpha^2)
    \frac{\dd\gamma(\alpha)}{\dd\alpha}.
 \end{equation*}
 Integrating this equation by parts, we get
 \begin{equation}\label{d1}
    d=\frac{3\pi}{8\mathcal{K}_\mathrm{F}}
     -\frac{3}{\mathcal{K}_\mathrm{F}^3}
     \int\limits_0^{\mathcal{K}_\mathrm{F}}\!\!
    \dd\alpha\,\alpha\,\gamma(\alpha).
 \end{equation}
 Taking into account that $\gamma(\alpha)=\arcsin\frac{\alpha}{s}$,
  we get
 \begin{equation}\label{d2}
    d=\frac{3\pi}{8\mathcal{K}_\mathrm{F}}
     -\frac{3}{4\mathcal{K}_\mathrm{F}}
     \left(
      \sqrt{\widetilde{s}^2-1}+
      \left(2-\widetilde{s}^2\right)
      \arcsin\frac{1}{\widetilde{s}}
     \right).
 \end{equation}
 Note that, if in Eq.~\eqref{d2} we put
 the magnitude of the Fermi wave vector $\mathcal{K}_\mathrm{F}^0$
 of noninteracting electrons,
 \begin{equation}\label{KF0}
  \mathcal{K}_\mathrm{F}^0=\left(\frac{9\pi}{4}\right)^{1/3}\frac1{r_\mathrm{s}},
 \end{equation}
 instead the magnitude of the Fermi wave vector $\mathcal{K}_\mathrm{F}$
 of interacting electrons,
 we get the well-known equation for noninteracting electrons~\cite{Kiejna,Huntington,Stratton}.

 \begin{widetext}

\begin{figure}[hbtp]
  \centering
  \includegraphics[width=0.49\textwidth]{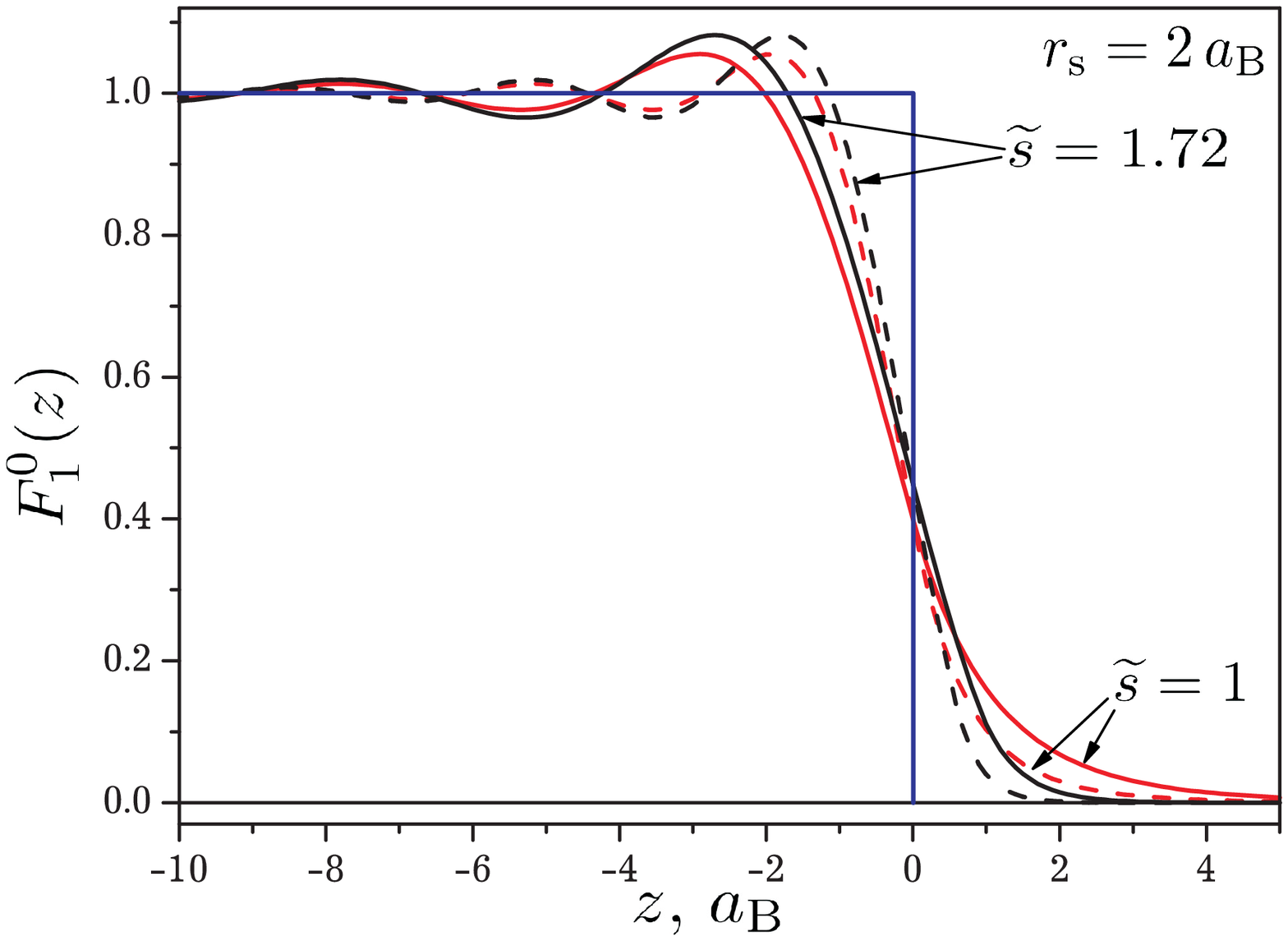}\hfill
  \includegraphics[width=0.49\textwidth]{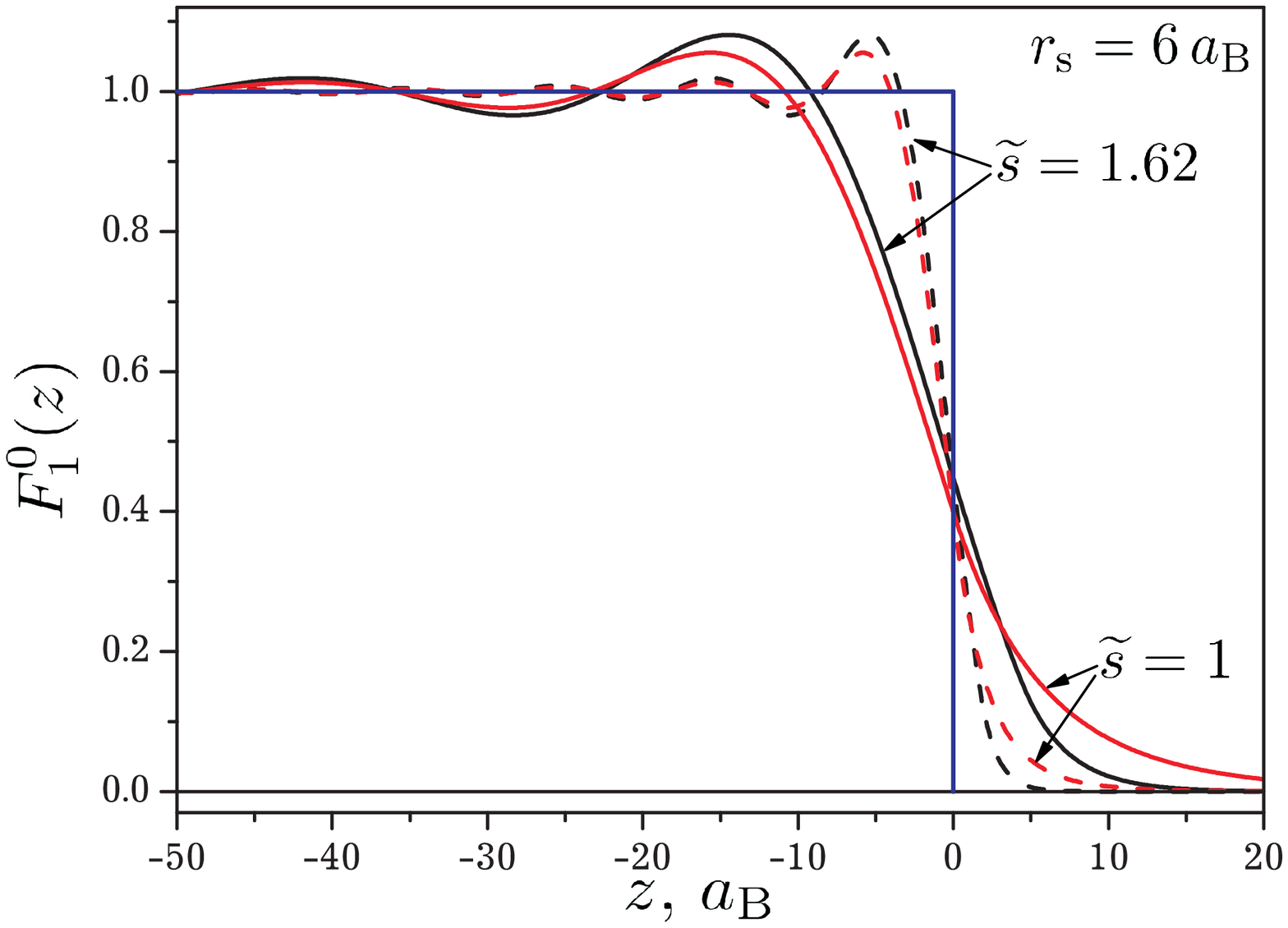}  \\
  \caption{The one-particle distribution function of electrons as
           a function of the electron coordinate normal
           to the dividing plane at $r_\mathrm{s}=2\,a_\mathrm{B}$ (left)
           and $r_\mathrm{s}=6\,a_\mathrm{B}$ (right)
           for different values of the barrier height parameter
           (the solid line is for interacting electrons whereas
           the dashed line is for noninteracting electrons).}\label{F01rs=2}
\end{figure}
%

\begin{figure}[hbtp]
  \centering
  \includegraphics[width=0.49\textwidth]{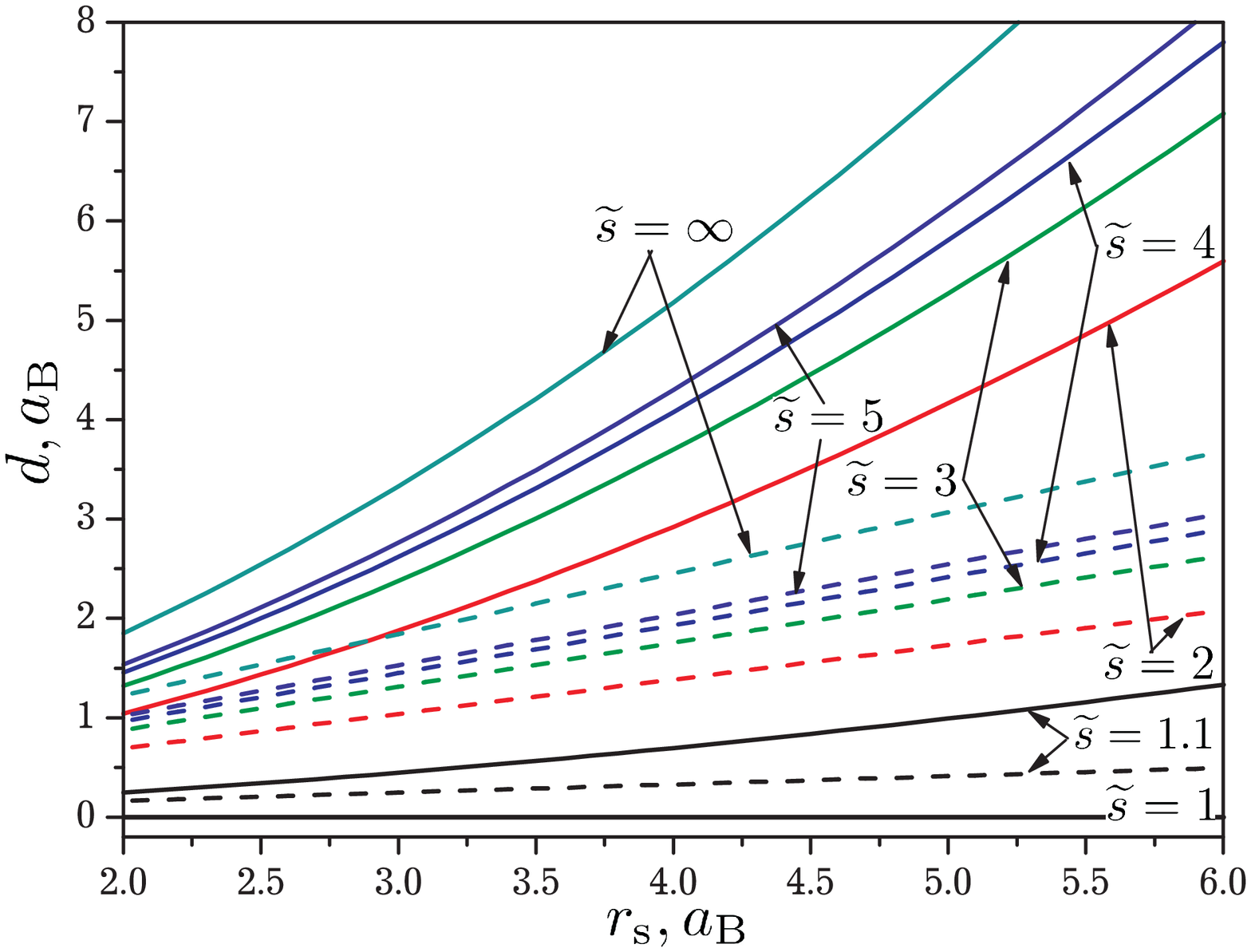}\hfill
 \includegraphics[width=0.49\textwidth]{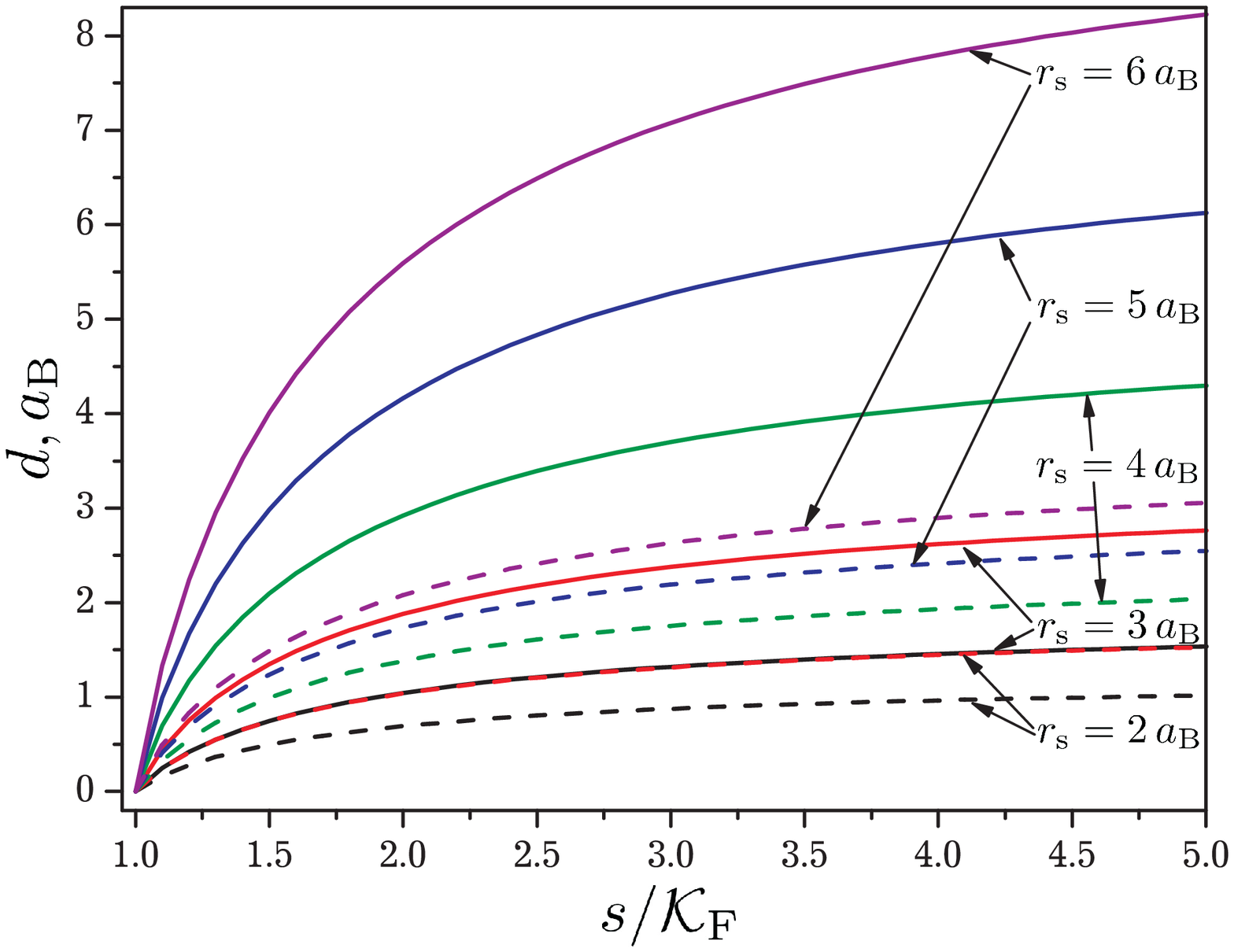}\\
  \caption{The parameter~$d$ as a function of the Wigner-Seitz radius at different
           values of the barrier height parameter (left)
           and as a function of the barrier height parameter~$s$
           at different values of the Wigner-Seitz radius (right)
           (the solid line is for interacting electrons whereas the dashed line is for noninteracting electrons).}
           \label{d_rs}
\end{figure}
 \end{widetext}

 In Fig.\,\ref{d_rs} (left),
 the parameter $d$~\eqref{d2} as a function of the Wigner-Seitz radius $r_\mathrm{s}$ is given
 for different values of the barrier height parameter of the step potential.
 Solid line represents the parameter~$d$ for interacting system,
 dashed line --- for noninteracting system.
 This parameter is the distance from the surface potential to the dividing plane.
 We see that taking into account the Coulomb interaction between electrons
 leads to an increase of this distance and its nonlinear dependence on $r_\mathrm{s}$,
 whereas the parameter $d$ for the noninteracting system is a linear function of $r_\mathrm{s}$.
 In the case of noninteracting electrons,
 this distance increases linearly with increasing of Wigner-Seitz radius,
 because the average distance between the electrons increases,
 and electrons can travel farther into the region~${z\geqslant0}$.
 The Coulomb repulsion between the electrons leads to
 an additional increase in the average distance between the electrons.
 Therefore, electrons can travel even farther into the region~${z\geqslant0}$,
 this distance as a function of Wigner-Seitz radius increases faster than linearly.

 In Fig.\,\ref{d_rs} (right),
 the parameter $d$~\eqref{d2} as a function of the barrier height parameter~$s$
 is given for different values of the Wigner-Seitz radius $r_\mathrm{s}$.
 For the barrier height~$W$,
 which is equal to the chemical potential~$\mu$,
 the distance from the dividing plane (${z=0}$)
 to the potential barrier ($ {z = d} $) is zero.
 Increase of the barrier height leads to an increase of the distance~$d$, and
 \[
  \lim_{s\to\infty}d=\frac{3\pi}{8\mathcal{K}_\mathrm{F}},
 \]
 that is the distance from the dividing plane
 to the infinite barrier model Refs.\,\cite{KMpreprint2014,KMprb2015}.

 \section{Internal energy}

 The internal energy of the system can be obtained from
 the thermodynamic potential $\Omega$ and the Gibbs-Helmholtz equation
 generalized for the case of variable number of particles,
 \begin{equation*}
    U=\Omega-\theta\frac{\partial\Omega}{\partial\theta}-\mu\frac{\partial\Omega}{\partial\mu}.
 \end{equation*}
 In the case of low temperatures (${\theta\to0}$),
 the second term of the right-hand side
 of this equation vanishes and we get
 \begin{equation}\label{U1}
    U=\Omega+\mu \langle N\rangle,
 \end{equation}
 where we have used the relation
 \begin{equation}\label{N}
    \langle N\rangle=-\frac{\partial\Omega}{\partial\mu}
 \end{equation}
 is the average number operator of electrons
 (the averaging, in contrast to~\eqref{N0},
 is performed with consideration of the Coulomb interaction between the electrons~\cite{KMpreprint2014,KMprb2015}).

 In Refs.~\cite{KMpreprint2014,KMprb2015}, it is shown that
  \begin{align}\label{N1}
    \langle N\rangle&=\sum\limits_{\mathbf{k}_{||},\alpha}n_\alpha(\mathbf{k}_{||})
    -\frac1{2S}\sum\limits_{\mathbf{q}\neq0}
                     \sum\limits_{\mathbf{k}_{||},\alpha}
                     \frac{\partial n_\alpha(\mathbf{k}_{||})}{\partial\mu}\!\!
      \int\limits_{-\frac L2}^{+\frac L2}\!\!\dd z\,|\varphi_\alpha(z)|^2 \nonumber\\
     &\qquad\times \big(g(\mathbf{q},z,z)-\nu(\mathbf{q},0)\big)\nonumber\\
    &\quad+\frac1{2S}\sum\limits_{\mathbf{q}\neq0}
                     \sum\limits_{\mathbf{k}_{||},\alpha_1,\alpha_2}
      \frac{\partial \big(n_{\alpha_1}(\mathbf{k}_{||}) \,
      n_{\alpha_2}(\mathbf{k}_{||}-\mathbf{q})\big)}{\partial\mu} \nonumber\\
    &\qquad\times\int\limits_{-\frac L2}^{+\frac L2}\!\!\dd z_1\!\!
      \int\limits_{-\frac L2}^{+\frac L2}\!\!\dd z_2\,
      \varphi^*_{\alpha_1}\!(z_1) \varphi^{\vphantom{*}}_{\alpha_2}\!(z_1)
      \varphi^*_{\alpha_2}\!(z_2) \varphi^{\vphantom{*}}_{\alpha_1}\!(z_2)\nonumber\\
      &\qquad\times g(\mathbf{q},z_1,z_2),
 \end{align}
 where $g(\mathbf{q},z_1,z_2)\equiv g(\mathbf{q},z_1,z_2,1)$ is
 the effective interelectron interaction in $(\mathbf{q},z)$ representation.

 Substituting Eqs.~\eqref{omega1} and~\eqref{N1} in Eq.~\eqref{U1},
 the internal energy~$U$ can be represented as
 \begin{equation}\label{U}
    U=U_0+\Delta U_1+\Delta U_2,
 \end{equation}
 where
 \begin{equation*}
    U_0=\Omega_0+\mu\sum\limits_{\mathbf{k}_{||},\alpha}n_\alpha(\mathbf{k}_{||})
 \end{equation*}
 is the internal energy of the noninteracting system
 (the calculation of~$U_0$ is done in Appendix~\ref{OmegaNonInteracting})
 though it indirectly takes into account the Coulomb interaction between electrons via
 the chemical potential $\mu$ of interacting electrons.
  \begin{widetext}
 \begin{align}\label{DeltaU1}
    \Delta U_1&=\frac1{2S}\sum\limits_{\mathbf{q}\neq0}
                     \sum\limits_{\mathbf{k}_{||},\alpha}
       \int\limits_{-\frac L2}^{+\frac L2}\!\!\!\dd z\,|\varphi_\alpha(z)|^2
      \Bigg[
           n_\alpha(\mathbf{k}_{||})\!\!
           \int\limits_0^1\!\!\dd\lambda \,\big(g(\mathbf{q},z,z,\lambda)-\nu(\mathbf{q},0)\big)- \mu\frac{\partial n_\alpha(\mathbf{k}_{||})}{\partial\mu}
       \big(g(\mathbf{q},z,z)-\nu(\mathbf{q},0)\big)
      \Bigg],
 \end{align}
 \begin{align}\label{DeltaU2}
    \Delta U_2&=\frac1{2S}\sum\limits_{\mathbf{q}\neq0}
                     \sum\limits_{\mathbf{k}_{||},\alpha_1,\alpha_2}
                     \int\limits_{-\frac L2}^{+\frac L2}\!\!\dd z_1\!\!
      \int\limits_{-\frac L2}^{+\frac L2}\!\!\dd z_2\,
      \varphi^*_{\alpha_1}\!(z_1) \varphi^{\vphantom{*}}_{\alpha_2}\!(z_1)
      \varphi^*_{\alpha_2}\!(z_2) \varphi^{\vphantom{*}}_{\alpha_1}\!(z_2)\nonumber\\
      &\quad\times\Bigg[\mu\frac{\partial \big(n_{\alpha_1}(\mathbf{k}_{||}) \,
      n_{\alpha_2}(\mathbf{k}_{||}-\mathbf{q})\big)}{\partial\mu}
      \,g(\mathbf{q},z_1,z_2)-
      n_{\alpha_1}(\mathbf{k}_{||}) \,
      n_{\alpha_2}(\mathbf{k}_{||}-\mathbf{q})
      \int\limits_0^1\!\!\dd\lambda \,g(\mathbf{q},z_1,z_2,\lambda)\Bigg].
 \end{align}
 The calculation of the sums
 of the Fermi-Dirac distribution
 over the wave vector
 in a plane parallel to the dividing plane $\mathbf{k}_{||}$
 at low temperature is done in Refs.~\cite{KMpreprint2014,KMprb2015}.
 Let us give here the results of these calculations.

 Thus,
 \begin{align*}
   &\sum\limits_{\mathbf{k}_{||}}n_\alpha(\mathbf{k}_{||})
     =\frac{S}{2\pi}\big(\mathcal{K}_\mathrm{F}^2-\alpha^2\big)\,\heav\big(\mathcal{K}_\mathrm{F}-\alpha\big),
   &\sum\limits_{\mathbf{k}_{||}}\frac{\partial n_\alpha(\mathbf{k}_{||})}{\partial\mu}
     =\frac{S}{2\pi}\frac{2m}{\hbar^2}\,\heav\big(\mathcal{K}_\mathrm{F}-\alpha\big),\\
 \end{align*}
 \begin{align*}
    &\sum\limits_{\mathbf{k}_{||}}n_{\alpha_1}(\mathbf{k}_{||})n_{\alpha_2}(\mathbf{k}_{||}-\mathbf{q})
    =\frac{2S}{(2\pi)^2}J(q,\alpha_1,\alpha_2),\nonumber\label{J1}
   & \sum\limits_{\mathbf{k}_{||}}
    \frac{\partial\big( n_{\alpha_1}(\mathbf{k}_{||})\,n_{\alpha_2}(\mathbf{k}_{||}-\mathbf{q})\big)}{\partial\mu}
    =\frac{2S}{(2\pi)^2}\frac{4m}{\hbar^2}J'(q,\alpha_1,\alpha_2),
 \end{align*}
 where
 \begin{equation*}
  J(q,\alpha_1,\alpha_2)=
    \left\{
      \begin{array}{cl}
        \left\{
         \begin{array}{ll}
            \pi c_1^2, & c_2>c_1, \\
            \pi c_2^2, & c_1>c_2,
         \end{array}
        \right. & 0\leqslant q<|c_1-c_2|,\\[4mm]
        f(c_1,c_2,q) + f(c_2,c_1,q),& |c_1-c_2|\leqslant q< c_1+c_2, \\[4mm]
        0,& q\geqslant c_1+c_2,
      \end{array}
    \right.
 \end{equation*}
 \begin{align*}
   f(c_1,c_2,q)=&\,\,c_1^2\left(\frac{\pi}{2}-\arcsin\frac{c_1^2-c_2^2+q^2}{2qc_1}\right)
   -\frac{c_1^2-c_2^2+q^2}{2q}\sqrt{c_1^2-\frac{(c_1^2-c_2^2+q^2)^2}{4q^2}},
 \end{align*}
 \begin{equation*}
   J'(q,\alpha_1,\alpha_2)=
    \left\{
      \begin{array}{cl}
        \left\{
         \begin{array}{ll}
            0, & 0\leqslant q\leqslant c_1-c_2, \\
            \arccos\dfrac{q^2+c_1^2-c_2^2}{2c_1q}, & c_1-c_2<q\leqslant c_1+c_2,\\
            0, & c_1+c_2<q<\infty,
         \end{array}
        \right\}, & c_1>c_2,\\[12mm]
        \left\{
         \begin{array}{ll}
            \pi, & 0\leqslant q\leqslant c_2-c_1, \\
            \arccos\dfrac{q^2+c_1^2-c_2^2}{2c_1q}, & c_2-c_1<q\leqslant c_1+c_2,\\
            0, & c_1+c_2<q<\infty,
         \end{array}
        \right\}, & c_2\geqslant c_1,
      \end{array}
    \right.
 \end{equation*}
  \[
  c_1=\sqrt{\mathcal{K}_\mathrm{F}^2-\alpha_1^2},\quad
  c_2=\sqrt{\mathcal{K}_\mathrm{F}^2-\alpha_2^2}.
 \]

 The calculated results for the integrals of products of the wave functions and
 the  effective potential are given in Appendix~\ref{intg}.
 \end{widetext}

 \section{Surface energy}

 Since the main aim of this work is to calculate of
 the free surface energy~$\sigma$,
 then it is necessary to single out the surface contribution~$U_\surf$
 (it is proportional to the area of the dividing plane~$S$)
 from the internal energy~\eqref{U}.
 Then the surface contribution to the internal energy per unit area of the dividing plane will be a required free surface energy, i.e.,
 \begin{align}\label{SE}
   \sigma&=\frac{U_\surf}{S}=\frac{U_{0,\surf}+\Delta U_{1,\surf}+\Delta U_{2,\surf}}{S}\nonumber\\
         &=\sigma_0+\Delta \sigma_1+\Delta \sigma_2,
 \end{align}
 where $U_{0,\surf} $ is the surface contribution to the internal energy of the noninteracting system
 (the calculation of $U_{0,\surf}$ is done in Appendix~\ref{OmegaNonInteracting},
 see Eq.~\eqref{U0surf}),
 \begin{align}\label{SE0}
    \sigma_{0}&=\frac{U_{0,\surf}}{S}\nonumber\\
    &=\frac{\hbar^2\mathcal{K}_\mathrm{F}^4}{160\pi m}
    \bigg[
     1+\frac1{2\pi}
     \bigg(
      \left(15\widetilde{s}^2-14\right)\sqrt{\widetilde{s}^2-1}\nonumber\\
     &\hspace*{20mm} -\left(15\widetilde{s}^4-24\widetilde{s}^2+8\right)
      \arcsin\frac{1}{\widetilde{s}}
     \bigg)
    \bigg]
 \end{align}
 is the surface energy of noninteracting system,
 \begin{align*}
    \Delta\sigma_1&=\frac{\Delta U_{1,\surf}}{S}\\
    &=\frac1{2\pi^2}a_\mathrm{B}^3
    \int\limits_0^\infty\!\!\dd q\,q\!
    \int\limits_0^{\mathcal{K}_\mathrm{F}}\!\!\dd \alpha
     \Bigg[
     (\mathcal{K}_\mathrm{F}^2-\alpha^2)
     \int\limits_0^1\!\!\dd\lambda\,
     \Delta G(q,\alpha,\lambda)\\
   &\qquad  -\mathcal{K}_\mathrm{F}^2\Delta G(q,\alpha,1)
    \Bigg]
    \frac{e^2}{a_\mathrm{B}^3},\\
    \Delta\sigma_2&=\frac{\Delta U_{2,\surf}}{S}\\
    &=\frac2{\pi^4}a_\mathrm{B}^3
    \int\limits_0^\infty\!\!\dd q\,q\!
    \int\limits_0^{\mathcal{K}_\mathrm{F}}\!\!\dd \alpha_1\!\!
    \int\limits_0^{\mathcal{K}_\mathrm{F}}\!\!\dd \alpha_2
    \Bigg[
     \mathcal{K}_\mathrm{F}^2J'(q,\alpha_1,\alpha_2)G(q,\alpha,1)\\
   &\qquad-\frac12J(q,\alpha_1,\alpha_2)
     \int\limits_0^1\!\!\dd\lambda\,
      G(q,\alpha,\lambda)
    \Bigg]
    \frac{e^2}{a_\mathrm{B}^3},
 \end{align*}
 where the transitions from the sums to the integrals are performed according to Eq.~\eqref{SumToIntAlpha}.
 Expressions for functions $\Delta G(q,\alpha,\lambda)$ and $G(q,\alpha,\lambda)$
 are given in Appendix~\ref{intg} (see Eqs.~\eqref{DeltaG} and \eqref{funkG} respectively).

 Note that, if in Eq.~\eqref{SE0} we put
 the magnitude of the Fermi wave vector $\mathcal{K}_\mathrm{F}^0$
 of noninteracting electrons Eq.~\eqref{KF0},
 instead of the magnitude of the Fermi wave vector $\mathcal{K}_\mathrm{F}$
 of interacting electrons,
 we get the well-known equation~\eqref{SE0} for the surface energy of noninteracting system~\cite{Kiejna,Huntington,Stratton}.

\begin{widetext}

\begin{figure}[hbtp]
  \centering
  \includegraphics[width=0.49\textwidth]{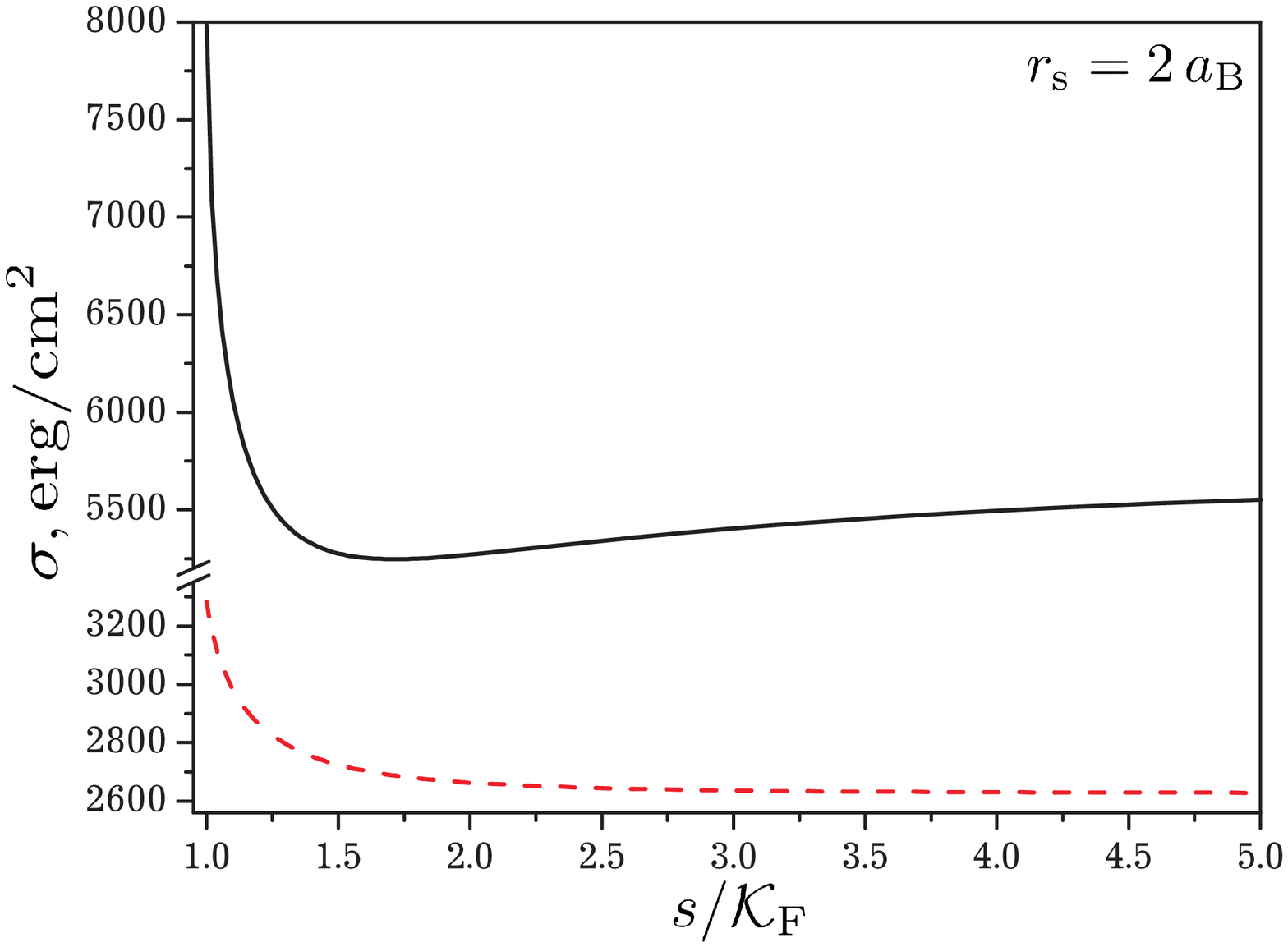}\hfill
  \includegraphics[width=0.49\textwidth]{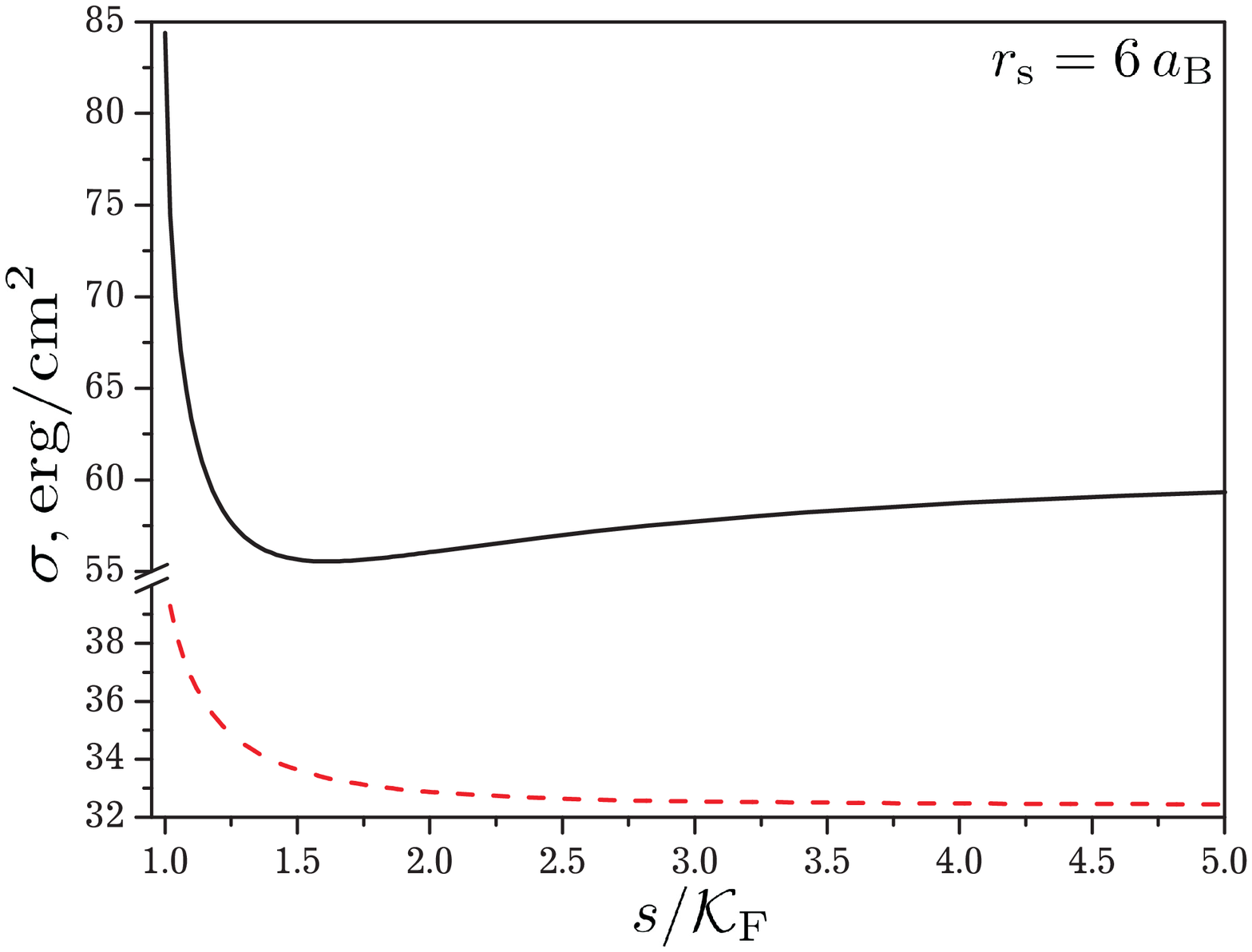}\\
  \caption{The surface energy as a function of the barrier height parameter~$s$ at $r_\mathrm{s}=2\,a_\mathrm{B}$ (left) and $r_\mathrm{s}=6\,a_\mathrm{B}$ (right)
           (the solid line is for interacting electrons whereas the dashed line is for noninteracting electrons).}\label{SE_s_rs=2}
\end{figure}
 \end{widetext}

 In Fig.\,\ref{SE_s_rs=2},
 the dependence of the surface energy~$\sigma$ on the barrier height parameter~$s$
 is presented for different values of the Wigner-Seitz radius.
 The solid line is for interacting electrons (see Eq.~\eqref{SE})
 whereas the dashed line is for interacting electrons (see Eq.~\eqref{SE0}).
 It can be concluded that
 if the barrier height of the step potential increases,
 the surface energy tends to the value,
 which is obtained for the infinite barrier model~\cite{KMpreprint2014,KMprb2015}.
 If the barrier height narrows down to the chemical potential,
 the surface energy of noninteracting system increases.
 It is clear,
 because in this case
 the average distance between the electrons increases,
 electrons can travel even farther into the region ${z\geqslant0}$,
 and therefore the surface energy increases.
 Taking into account the Coulomb interaction between electrons
 leads to a significant increase in the surface energy,
 its dependence on the barrier height parameter~$s$ is no longer monotonic,
 and the surface energy as a function of the parameter~$s$ has a minimum.
 Since a system always tends to the lowest energy state,
 the minimum of the surface energy can be seen
 as self-consistent condition for the barrier height of the step potential
 (the values of the parameter~$s$ are presented in Tab.\,\ref{table1}
 for different values of the Wigner-Seitz radius).

\begin{table}[h!]
  \centering
  \caption{ The values of the surface energy minimum and its coordinates.}\label{table1}
  \begin{tabular}{c||c|c|c|c|c}
    \hline
    \rule{0mm}{2mm}$r_\mathrm{s},\,a_\mathrm{B}$ & 2 & 3 & 4 & 5 & 6 \\ \hline
    \rule{0mm}{3.5mm}${s}/\mathcal{K}_\mathrm{F}$ & $1.72$ & $1.66$ & $1.64$ & $1.63$ & $1.62$ \\ \hline
    \rule{0mm}{3.5mm}$\sigma,\,\mathrm{erg}/\mathrm{cm}^2$ & 5246 & 1067 & 322 & 123 & 55 \\
    \hline
  \end{tabular}
\end{table}

\begin{figure}
  \centering
  \includegraphics[width=0.45\textwidth]{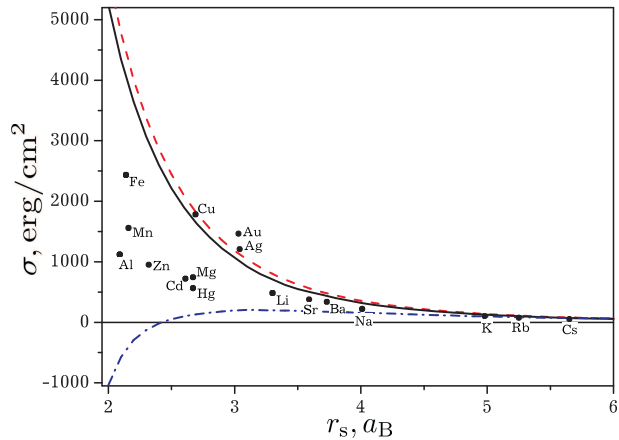}\\
  \caption{The surface energy as a function of the Wigner-Seitz radius
           (the solid line is for the height of the potential barrier,
           which fulfills the condition for minimum of the surface energy;
           the dashed line is for the infinite barrier model~\cite{KMpreprint2014,KMprb2015},
           the dash-dotted line is the result of Lang and Kohn~\cite{Lang};
           whereas the dots are experimental data~\cite{Vitos}.\label{SE_rs}}
\end{figure}

 In Fig.\,\ref{SE_rs},
 the dependence of the surface energy~$\sigma$ on the Wigner-Seitz radius~$r_\mathrm{s}$ is presented.
 The solid line is the surface energy calculated for the values of the barrier height parameter~$s$
 fulfilled the condition for minimum of the surface energy.
 The dashed line is the surface energy for the infinite barrier model~\cite{KMpreprint2014,KMprb2015},
 the dash-dotted line is the well-known result of Lang and Kohn~\cite{Lang},
 and the dots are experimental data for some metals according to Ref.~\cite{Vitos}.

 The results given in this figure show that
 the calculated values of the surface energy for the step potential model is positive
 in the entire domain of the Wigner-Seitz~$r_\mathrm{s}$,
 these values are lower than the values of the surface energy for the infinite barrier model,
 and in the domain~$r_\mathrm{s}>5a_\mathrm{B}$,
 the values of the surface energy for finite and infinite barrier model
 are in good agreement with the well-known result of Lang and Kohn~\cite{Lang}.
 In addition,
 for such a simple model of semi-bounded metal,
 which is the semi-infinite jellium,
 the calculated values of the surface energy are in sufficiently good agreement
 with experimental data for some metals.
 Obviously,
 incorporation of discreteness of ionic subsystem is necessary to better agreement
 with experimental data.

 \section{Conclusions}

 By using the step potential model for the surface potential,
 the one-particle distribution function of electrons,
 the distance from the surface potential to the dividing plane,
 and the surface energy of the semi-bounded metal
 within the framework of the semi-infinite jellium
 are calculated and studied at low temperatures.

 It is found that the
 taking into account the Coulomb interaction between electrons
 leads to an increase in the period of damped oscillations around its average value in the bulk of the metal,
 and increasing of the barrier height of the step potential leads to more rapid damping
 of the one-particle distribution function near the dividing plane.

 It is shown that the taking into account the Coulomb interaction between electrons
 leads to an increase in the distance between the dividing plane and
 the surface potential,
 and its nonlinear dependence on the Wigner-Seitz radius,
 whereas this distance for the noninteracting system is a linear function.
 The Coulomb repulsion between the electrons leads to
 an additional increase of the average distance between the electrons.
 Therefore electrons can travel even farther into the region~${z\geqslant0}$,
 this distance as a function of Wigner-Seitz radius increases faster than linearly.

 It is found that
 taking into account the Coulomb interaction between electrons
 leads to a significant increase in the surface energy,
 its dependence on the barrier height of the step potential is no longer monotonic,
 whereas the surface energy of the noninteracting system
 is monotonically decreasing function.
 There is the minimum of the surface energy at some value of the barrier height.
 The condition of this minimum is used as a self-consistent condition
 for the barrier height at different values of the Wigner-Seitz radius.
 The obtained values of the barrier height of the step potential
 decrease with increasing of the Wigner-Seitz radius.
 Using these values,
 the surface energy is calculated as a function of the Wigner-Seitz radius,
 and it is lower than the surface energy for the infinite barrier model of the surface potential Refs.\,\cite{KMpreprint2014,KMprb2015}.

 In contrast to the surface energy calculated by Lang  and Kohn,
 the surface energy of semi-bounded metal within the framework of the semi-infinite jellium
 calculated by us is positive in the entire area of the Wigner-Seitz radius,
 and it is in sufficiently good agreement with experimental data.

 \appendix
  \renewcommand{\theequation}{\Alph{section}.\arabic{equation}}

 \section{Thermodynamical potential and internal  energy of noninteracting system}\label{OmegaNonInteracting}

 Let us calculate the thermodynamical potential of noninteracting system,
 \begin{equation*}
 \Omega_0=-\frac1\beta\sum\limits_{\mathbf{k}_{||},\alpha}
 \ln\left[1+\ee^{\beta(\mu-E_\alpha(\mathbf{k}_{||}))}\right].
 \end{equation*}
 Since here $\mu$ is the chemical potential of interacting electron system,
 the Coulomb interaction in this expression is taken into account
 indirectly via chemical potential.

 To perform summation with respect to $\mathbf{k}_{||}$ and $\alpha$,
  we use the density of states Ref.\,\cite{KMpreprint2014},
 \begin{align}\label{DOSFBM}
    \rho(E)=&\frac{SL}2\frac{\sqrt{2}m^{3/2}}{\pi^2\hbar^3}\sqrt{E}+\nonumber\\
     &+S\left[\frac{\sqrt{2}m^{3/2}d}{\pi^2\hbar^3}\sqrt{E}
      +\frac{m}{\pi^2\hbar^2}\,\gamma\!\left(\textstyle\frac{\sqrt{2mE}}{\hbar}\right)
      -\frac{m}{4\pi\hbar^2}\right].
 \end{align}
 At low temperatures (${\beta\to\infty}$),
 the thermodynamical potential of noninteracting system has the form
 \begin{equation}\label{Omega0Suma}
  \Omega_0=\Omega_{0,\bulk}+\Omega_{0,\surf},
 \end{equation}
 where
 \begin{align}\label{omega2bulk}
   \Omega_{0,\bulk}=&-\frac{SL}2\frac{4\sqrt{2}m^{3/2}}{15\pi^2\hbar^3}\mu^{5/2}=\nonumber\\
                   =&-\frac{SL}2\frac{\hbar^2}{15m\pi^2}\mathcal{K}_\mathrm{F}^5
 \end{align}
 is the extensive contribution to the thermodynamical potential of noninteracting system
 (it is proportional to the volume of the system ${SL}$),
 which is dependent on the magnitude of the Fermi wave vector $\mathcal{K}_\mathrm{F}$
 of interacting system of electrons,
 \begin{align}\label{omega2surf}
   \Omega_{0,\surf}&=-S\Bigg[\frac{4\sqrt{2}m^{3/2}d}{15\pi^2\hbar^3}\mu^{5/2}+\nonumber\\
         &\quad+\frac{m}{\pi^2\hbar^2}\int\limits_0^\mu\!\dd E\,
          (\mu-E)\,\gamma\!\left(\textstyle\frac{\sqrt{2mE}}{\hbar}\right)
         -\frac{m}{8\pi\hbar^2}\mu^2\Bigg]\nonumber\\
   &=S\,\frac{\hbar^2\mathcal{K}_\mathrm{F}^4}{m\pi^2}
           \Bigg[\frac{\pi}{32}-\frac{d\,\mathcal{K}_\mathrm{F}}{15}\nonumber\\
   &\qquad      -\frac{1}{2\mathcal{K}_\mathrm{F}^4}
            \int\limits_0^{\mathcal{K}_\mathrm{F}}\!\!\dd \alpha\,\alpha\,(\mathcal{K}_\mathrm{F}^2-\alpha^2)\,
            \gamma(\alpha)\Bigg]
 \end{align}
 is the surface contribution
 (it is proportional to the area of the dividing plane $S$).
 Taking into account Eq.~\eqref{d1} for the parameter $d$,
 we get that
 \begin{equation*}
    \Omega_{0,\surf}=S\frac{\hbar^2\mathcal{K}_\mathrm{F}^4}{160\pi m}\!
    \left[
     1+
     \frac{80}{\pi\mathcal{K}_\mathrm{F}^4}\!\!
     \int\limits_0^{\mathcal{K}_\mathrm{F}}\!\!
     \dd\alpha\,\alpha
     \left(\alpha^2-\frac35\mathcal{K}_\mathrm{F}^2\right)
     \gamma(\alpha)
    \right].
 \end{equation*}

 Using Eqs.~\eqref{Omega0Suma}--\eqref{omega2surf},
 the average of the number operator of electrons
 without taking into account the Coulomb interaction between electrons
 can be represented as
 \begin{equation*}
    \langle N\rangle_0=-\frac{\partial\Omega_0}{\partial\mu}=N_{0,\bulk}+N_{0,\surf},
 \end{equation*}
 where
 \begin{equation*}
    N_{0,\bulk}=-\frac{\partial\Omega_{0,\bulk}}{\partial\mu}
    =\frac{SL}{2}\frac{\mathcal{K}_\mathrm{F}^3}{3\pi^2},
 \end{equation*}
 \begin{align}\label{N0surf}
    N_{0,\surf}&=-\frac{\partial\Omega_{0,\surf}}{\partial\mu}
    =\nonumber\\
    &=S\Bigg[\frac{2\sqrt{2}m^{3/2}d}{3\pi^2\hbar^3}\mu^{3/2} \nonumber\\
    &\quad+\frac{m}{\pi^2\hbar^2}\int\limits_0^\mu\!\dd E\,
          \gamma\!\left(\textstyle\frac{\sqrt{2mE}}{\hbar}\right)
         -\frac{m}{4\pi\hbar^2}\mu\Bigg]=\nonumber\\
    &=S\frac{\mathcal{K}_\mathrm{F}^2}{\pi^2}
      \left[
       \frac{\mathcal{K}_\mathrm{F}d}{3}-\frac\pi8+
       \frac1{\mathcal{K}_\mathrm{F}^2}
       \int\limits_0^{\mathcal{K}_\mathrm{F}}\!\!\dd\alpha\,\alpha\,\gamma(\alpha)
      \right]   .
 \end{align}
 Taking into account Eq.~\eqref{d1} for the parameter $d$,
 we get that
 \begin{equation}\label{N0surf1}
    N_{0,\surf}=0.
 \end{equation}

 At low temperatures,
 the internal energy of noninteracting system
 can be represented as
 \begin{align*}
    U_0&=\Omega_0+\mu\langle N\rangle_0
       =U_{0,\bulk}+U_{0,\surf},
 \end{align*}
 where
 \begin{align*}
    U_{0,\bulk}&=\Omega_{0,\bulk}+\mu N_{0,\bulk}
      =\frac{SL}{2}\frac{\hbar^2\mathcal{K}_\mathrm{F}^2}{10\pi^2m},
 \end{align*}
 \begin{align*}
    U_{0,\surf}&=\Omega_{0,\surf}+\mu N_{0,\surf}=\Omega_{0,\surf}=\\
    &=S\frac{\hbar^2\mathcal{K}_\mathrm{F}^4}{160\pi m}\!
    \left[
     1+
     \frac{80}{\pi\mathcal{K}_\mathrm{F}^4}\!\!
     \int\limits_0^{\mathcal{K}_\mathrm{F}}\!\!
     \dd\alpha\,\alpha
     \left(\alpha^2-\frac35\mathcal{K}_\mathrm{F}^2\right)
     \gamma(\alpha)
    \right].
 \end{align*}
 Taking into account that $\gamma(\alpha)=\arcsin\frac{\alpha}{s}$,
  we get
 \begin{align}\label{U0surf}
    U_{0,\surf}&=S\frac{\hbar^2\mathcal{K}_\mathrm{F}^4}{160\pi m}
    \Bigg[
     1+\frac1{2\pi}
     \bigg(
      \left(15\widetilde{s}^2-14\right)\sqrt{\widetilde{s}^2-1}-\nonumber\\
     &\qquad-\left(15\widetilde{s}^4-24\widetilde{s}^2+8\right)
      \arcsin\frac{1}{\widetilde{s}}
     \bigg)
    \Bigg].
 \end{align}

  \section{The calculation of integrals with the effective interelectron interaction}
 \label{intg}

 In this Appendix the results of calculation of the integrals
 \begin{equation}\label{gphi2}
   \int\limits_{-\frac L2}^{+\frac L2}\!\!\dd z\,|\varphi_\alpha(z)|^2
     \big(g(\mathbf{q},z,z,\lambda)-\nu(\mathbf{q},0)\big),
 \end{equation}
 and
 \begin{equation}\label{gphi4}
   \int\limits_{-\frac L2}^{+\frac L2}\!\!\dd z_1\!\!
      \int\limits_{-\frac L2}^{+\frac L2}\!\!\dd z_2\,
      \varphi^*_{\alpha_1}\!(z_1) \varphi^{\vphantom{*}}_{\alpha_2}\!(z_1)
      \varphi^*_{\alpha_2}\!(z_2) \varphi^{\vphantom{*}}_{\alpha_1}\!(z_2)
      g(\mathbf{q},z_1,z_2,\lambda)
 \end{equation}
 are given.
 Here $\varphi_\alpha(z)$ are the wave function~\eqref{waveFunction}
 of electrons in the field of the step potential,
 which is located at the point ${z=d}$;
 $g(\mathbf{q},z_1,z_2,\lambda)$ is the effective interelectron interaction,
 which is a solution of the integral equation~\eqref{intEq1}
 and obtained using the technique of Refs.~\cite{JPS2003_1,CMP2006}.
 This potential depends on module of the vector~$\mathbf{q}$:
  \begin{widetext}\vspace{-2mm}
 \begin{align*}
   g({q},z_1\leqslant d,z_2\leqslant d,\lambda)&
   =\frac{2\pi e^2}{Q_1(\lambda)}
     \left[
      \ee^{-Q_1(\lambda)|z_1-z_2|}+
       \frac{Q_1(\lambda)-Q_2(\lambda)}{Q_1(\lambda)+Q_2(\lambda)}\ee^{Q_1(\lambda)(z_1+z_2-2d)}
     \right],\\
  g({q},z_1\geqslant d,z_2\geqslant d,\lambda)&
  =\frac{2\pi e^2}{Q_2(\lambda)}
    \left[
     \ee^{-Q_2(\lambda)|z_1-z_2|}-
      \frac{Q_1(\lambda)-Q_2(\lambda)}{Q_1(\lambda)+Q_2(\lambda)}\ee^{-Q_2(\lambda)(z_1+z_2-2d)}
    \right],\\
  g({q},z_1\geqslant d,z_2\leqslant d,\lambda)&=\frac{4\pi e^2}{Q_1(\lambda)+Q_2(\lambda)}
                        \,\ee^{Q_1(\lambda)(z_2-d)-Q_2(\lambda)(z_1-d)},\\
  g({q},z_1\leqslant d,z_2\geqslant d,\lambda)&=\frac{4\pi e^2}{Q_1(\lambda)+Q_2(\lambda)}
                        \,\ee^{Q_1(\lambda)(z_1-d)-Q_2(\lambda)(z_2-d)},
 \end{align*}
 where
 \begin{align*}
    Q_1(\lambda)&=\sqrt {q^2+\lambda\,\varkappa_{\mathrm{TF}}^2
  \big(\lindhard\big(\textstyle\frac{q}{2\mathcal{K}_{\mathrm{F}}}\big)
  -\Delta\big(\textstyle\frac{q}{2\mathcal{K}_{\mathrm{F}}}\big)\big)},
  &Q_2(\lambda)=\sqrt{q^2+\lambda\,\varkappa_{\mathrm{TF}}^2
  \Delta\big(\textstyle\frac{q}{2\mathcal{K}_{\mathrm{F}}}\big)},
 \end{align*}
 \[
  \Delta(x)=\frac{2}{\widetilde{s}^2}
  \int\limits_0^1\!\!\dd\xi\,\xi
  \sqrt{\widetilde{s}^2-\xi^2}
  \left[
   1-\sqrt{1-\frac{1-\xi^2}{x^2}}\,
   \heav\!\left(1-\frac{1-\xi^2}{x^2}\right)
  \right],\;
  \widetilde{s}=\frac{s}{\mathcal{K}_\mathrm{F}}.
 \]

 Note that the integrals~ \eqref{gphi2} and \eqref{gphi4} are equal to zero for~${\alpha=0}$,
 because in this case the wave functions~\eqref{waveFunction} are equal to zero.

 After integration, Eq.~\eqref{gphi2} for ${\alpha\neq0}$  has the form
 \begin{align*}
    \int\limits_{-\frac L2}^{+\frac L2}\!\!\dd z&\,|\varphi_\alpha(z)|^2
     \big(g({q},z,z,\lambda)-\nu(q,0)\big)=
     2\pi e^2|C(\alpha)|^2\frac{L}4
      \left(\frac{1}{Q_1(\lambda)}-\frac1q\right)
      +2\pi e^2|C(\alpha)|^2\Delta G(q,\alpha,\lambda),
 \end{align*}
 where
 \begin{align}\label{DeltaG}
    \Delta G(q,\alpha,\lambda)&=\left(\frac{d}2+\frac{\sin(2\gamma(\alpha))}{4\alpha}\right)
                        \left(\frac{1}{Q_1(\lambda)}-\frac1q\right)
                          +\frac1{2\varkappa(\alpha)}\left(\frac\alpha s\right)^2
              \left(\frac{1}{Q_2(\lambda)}-\frac1q\right)+\nonumber\\
    &\quad  +\frac1{4Q_1(\lambda)}\frac{Q_1(\lambda)-Q_2(\lambda)}{Q_1(\lambda)+Q_2(\lambda)}
     \left(\frac1{Q_1(\lambda)}-\frac{Q_1(\lambda)\cos(2\gamma(\alpha))-\alpha\sin(2\gamma(\alpha))}{Q_1^2(\lambda)+\alpha^2}\right)-\nonumber\\
    &\quad-\frac1{2Q_2(\lambda)} \left(\frac\alpha s\right)^2
     \frac{Q_1(\lambda)-Q_2(\lambda)}{Q_1(\lambda)+Q_2(\lambda)}
     \frac1{Q_2(\lambda)+\varkappa(\alpha)}.
 \end{align}

 After integration, Eq.~\eqref{gphi4} for ${\alpha_1\neq0}$ and ${\alpha_2\neq0}$ has the form
 \begin{align*}
    \int\limits_{-\frac L2}^{+\frac L2}&\!\dd z_1\!\!
      \int\limits_{-\frac L2}^{+\frac L2}\!\!\dd z_2\,
      \varphi^*_{\alpha_1}\!(z_1) \varphi^{\vphantom{*}}_{\alpha_2}\!(z_1)
      \varphi^*_{\alpha_2}\!(z_2) \varphi^{\vphantom{*}}_{\alpha_1}\!(z_2)
      g({q},z_1,z_2,\lambda)= \\
    &=2\pi e^2|C(\alpha_1)|^2|C(\alpha_2)|^2\frac{L}{4}
    \frac{Q_1^2(\lambda)+\alpha_1^2+\alpha_2^2}{\left(Q_1^2(\lambda)+\alpha_1^2+\alpha_2^2\right)^2-4\alpha_1^2\alpha_2^2}
   +2\pi e^2 |C(\alpha_1)|^2|C(\alpha_2)|^2 G(q,\alpha_1,\alpha_2,\lambda)\nonumber,
 \end{align*}
 where
 \begin{align}\label{funkG}
    G&(q,\alpha_1,\alpha_2,\lambda)=\frac14\big(f_1(\alpha_1,\alpha_2,\lambda)+f_1(\alpha_1,-\alpha_2,\lambda)\big)+\frac1{4Q_1(\lambda)}\frac{Q_1(\lambda)-Q_2(\lambda)}{Q_1(\lambda)+Q_2(\lambda)}
     \big(f_2(\alpha_1,\alpha_2,\lambda)-f_2(\alpha_1,-\alpha_2,\lambda)\big)^2\nonumber\\
    &+\frac{(\alpha_1\alpha_2)^2}{s^4}
           \frac1{Q_2(\lambda)}
            \left(
            \frac1{(\varkappa(\alpha_1)+\varkappa(\alpha_2))(Q_2(\lambda)+\varkappa(\alpha_1)+\varkappa(\alpha_2))}
           -\frac{Q_1(\lambda)-Q_2(\lambda)}{Q_1(\lambda)+Q_2(\lambda)}
             \frac1{(Q_2(\lambda)+\varkappa(\alpha_1)+\varkappa(\alpha_2))^2}
           \right)\nonumber\\
   &+\frac{\alpha_1\alpha_2}{s^2}
          \frac2{Q_1(\lambda)+Q_2(\lambda)}
          \frac1{Q_2(\lambda)+\varkappa(\alpha_1)+\varkappa(\alpha_2)}    \big(f_2(\alpha_1,\alpha_2,\lambda)-f_2(\alpha_1,-\alpha_2,\lambda)\big),
 \end{align}
 \begin{align*}
   f_1(\alpha_1,\alpha_2,\lambda)&=\frac1{Q_1^2(\lambda)+(\alpha_1-\alpha_2)^2}
        \Bigg[
         d-\frac{\sin(2(\gamma(\alpha_1)-\gamma(\alpha_2)))}{2(\alpha_1-\alpha_2)}
         +\frac{\sin(2\gamma(\alpha_1))}{2\alpha_1}
       +\frac{\sin(2\gamma(\alpha_2))}{2\alpha_2}
       +\frac{Q_1(\lambda)}{Q_1^2(\lambda)+(\alpha_1+\alpha_2)^2}\\
      &\quad-\frac{Q_1^2(\lambda)\big(1+\cos^2(\gamma(\alpha_1)-\gamma(\alpha_2))\big)
        -(\alpha_1-\alpha_2)^2\sin^2(\gamma(\alpha_1)-\gamma(\alpha_2))}
        {Q_1(\lambda)\big(Q_1^2(\lambda)+(\alpha_1-\alpha_2)^2\big)} \\
      &\quad+\frac{Q_1(\lambda)\cos(\gamma(\alpha_1)-\gamma(\alpha_2))+
        (\alpha_1-\alpha_2)\sin(\gamma(\alpha_1)-\gamma(\alpha_2))}
        {Q_1(\lambda)\big(Q_1^2(\lambda)+(\alpha_1+\alpha_2)^2\big)}\\
      &\qquad\times   \big(Q_1(\lambda)\cos(\gamma(\alpha_1)+\gamma(\alpha_2))-
        (\alpha_1+\alpha_2)\sin(\gamma(\alpha_1)+\gamma(\alpha_2))\big)  \Bigg],
 \end{align*}
 \begin{equation*}
  f_2(\alpha_1,\alpha_2,\lambda)
  =\frac{Q_1(\lambda)\cos(\gamma(\alpha_1)-\gamma(\alpha_2))-
        (\alpha_1-\alpha_2)\sin(\gamma(\alpha_1)-\gamma(\alpha_2))}
        {Q_1^2(\lambda)+(\alpha_1-\alpha_2)^2}.
 \end{equation*}
 \newpage\end{widetext}


\begin{thebibliography}{99}
\bibitem{footnote} Because the form of the thermodynamic potential~$\Omega_0$ as a function of $\mu$
   coincides with thermodynamic potential of an ideal electron gas,
   the thermodynamic potential~$\Omega_0$ is called by us
   thermodynamic potential of the noninteracting system though
   it indirectly takes into account the Coulomb interaction between electrons
   via the chemical potential~$\mu$ of interacting electrons.
   The same applies to the average of the number operator of electrons~$\langle N\rangle_0$,
   the internal energy~$U_0$,
   the surface energy~$\sigma_0$.
\end{thebibliography}
\end{document}